\documentclass[aip,jap,preprint,superscriptaddress]{revtex4-1}
\usepackage[utf8]{inputenc}
\usepackage{graphicx}
\graphicspath{ {./figures/} }

\begin{document}

\title{Interface-driven magnetic anisotropy in relaxed La$_{0.7}$Sr$_{0.3}$CrO$_3$/La$_{0.7}$Sr$_{0.3}$MnO$_3$ heterostructures on MgO}

\author{Sanaz Koohfar}
\author{Yasemin Ozbek}
\author{Hayden Bland}
\affiliation{Department of Physics, North Carolina State University, Raleigh, NC 27695, USA}
\author{Zhan Zhang}
\affiliation{Advanced Photon Source, Lemont, IL 76019, USA}

\author{Divine P. Kumah}
\email{Author to whom correspondence should be addressed: dpkumah@ncsu.edu}
\affiliation{Department of Physics, North Carolina State University, Raleigh, NC 27695, USA}

\begin{abstract}
We investigate the structural and magnetic properties of La$_{0.7}$Sr$_{0.3}$CrO$_3$ (LSCO)/La$_{0.7}$Sr$_{0.3}$MnO$_3$(LSMO)  heterostructures grown on (001)-oriented MgO by molecular beam epitaxy. Due to the large film-substrate lattice mismatch, strain relaxation is found to occur within the first 2-3 unit cells (uc) of the film as evidenced by reflection high energy electron diffraction and high-resolution synchrotron X-ray reciprocal space mapping. We find that the presence of the LSCO spacer and capping layers leads to ferromagnetism in ultra-thin LSMO layers with thicknesses on the order of 2 uc with the magnetic easy axis oriented in the film plane. Net magnetic moments of 1.4 and 2.4 $\mu_B$/Mn are measured for [2 uc  LSCO/ 2 uc LSMO] and [2 uc LSCO/ 4 uc LSMO] superlattices, respectively by SQUID magnetometry.  The effective magnetic anisotropy of the relaxed [2 uc LSCO/ 4 uc LSMO] heterostructure is found to be an order of magnitude higher than bulk LSMO highlighting the critical role of interfacial magnetic exchange interactions in tuning magnetic anisotropy at complex oxide interfaces.

\end{abstract}

\maketitle

\section{Introduction}

Doped rare-earth manganites (A$_{1-x}$B$_x$MnO$_3$ where A is a rare-earth ion and B is an alkaline-earth ion) exhibit a wide range of interesting physical properties including tunable magnetic phases and metal-insulator transitions, colossal magnetoresitivity and half metallicity.\cite{mott1964electrons, tokura2000colossal} For bulk La$_{1-x}$Sr$_x$MnO$_3$ (LSMO) with 30\% Sr doping, a ferromagnetic metallic state exists below 360 K. \cite{ikeda2010perpendicular,mangin2006current} Due to the strong coupling of the lattice, spin and electronic  degrees of freedom, the transport and magnetic properties of LSMO thin films have been tuned by epitaxial growth on closely lattice-matched single crystal substrates such as LaAlO$_3$, SrTiO$_3$, DyScO$_3$ and LSAT.\cite{tsui2000strain, Koohfar2019, konishi1999orbital} Epitaxial strain provides an effective route to control magnetic anisotropy in LSMO thin films with important implications for the design of novel spin-based devices.\cite{suzuki1997role, berndt2000magnetic, boschker2009strong, liao2016} The magnetic easy axis for LSMO films under tensile strain on SrTiO$_3$ lies in-plane along the [110] pseudocubic axis while compressively strained films on LaAlO$_3$ exhibit strong perpendicular magnetic anisotropy.\cite{suzuki1997role} As the film thickness is reduced to the dimensions on the order of a unit cell, surface and interfacial contributions to magnetic anisotropy energy are non-negligible and compete with strain-mediated contributions to magnetocrystalline anisotropy.\cite{Yi2016Atomic, song2020electric}

For uncapped stoichiometric LSMO films, below a critical thickness of 4-10 nm, ferromagnetism is suppressed limiting potential applications in thin film devices and a direct decoupling of the various contributions to magnetic anisotropy.\cite{tsui2000strain,Huijben2008CriticalFilms} The suppression of ferromagnetism in thin LSMO films has been attributed to structural distortions and chemical and electronic reconstructions arising from interfacial polar discontinuities,\cite{Peng2014, Koohfar2017, boschker2012preventing, Mundy2014}, oxygen vacancies\cite{Li2012}, cation disorder\cite{jin2016direct}, orbital reconstructions\cite{tebano2008evidence} and distortions of the oxygen octahedra due to interfacial structural coupling.\cite{huijben2017Interface, yi2017engineering, Bhattacharya2014MagneticHeterostructures, Moon2017StructuralHeterostructures} Recent reports on coherently strained LSMO layers in LSMO/SrRuO$_3$ superlattices\cite{ziese2012stabilization} and LSMO layers capped with La$_{0.7}$Sr$_{0.3}$CrO$_3$ (LSCO) have evidenced ferromagnetism in LSMO layers as thin as 2 unit cells (0.8 nm).\cite{Koohfar2019, koohfar2020effect, olmos2020exchange, penn2020redistribution} 

In this letter, we investigate the structural and magnetic properties of strain-free LSMO/LSCO heterostructures grown on (001)-oriented single crystal MgO substrates.\cite{gommert1999influence,casanove2002growth, Borges2001MagneticFilms} Bulk LSMO and LSCO have pseudocubic lattice constants of 3.88 and 3.86 \AA{} respectively. The lattice mismatch of LSMO and LSCO with cubic MgO (c=4.212 \AA{}) is 8.5\%. Due to the large tensile lattice mismatch between the MgO substrate and the LSMO and LSCO layers, strain relaxation is found to occur within the first 2-3 unit cells of the film as evidenced by \textit{in-situ} reflection high energy electron diffraction measurements and \textit{ex-situ} X-ray diffraction. A paramagnetic-ferromagnetic transition is observed for heterostructures with 2 unit cell (u.c.) thick LSMO layers indicative of the removal of magnetically dead LSMO layers. The relaxed states of the ultra-thin LSMO films allows us to elucidate the role of confinement to magnetic anisotropy in this system. The magnetic easy axis is found to lie in the plane of the film due to the dominant contributions of LSMO/LSCO interfacial exchange interactions and confinement to the total magnetic anisotropy.

\begin{figure}[t!]
    \centering
    \includegraphics[width=3in]{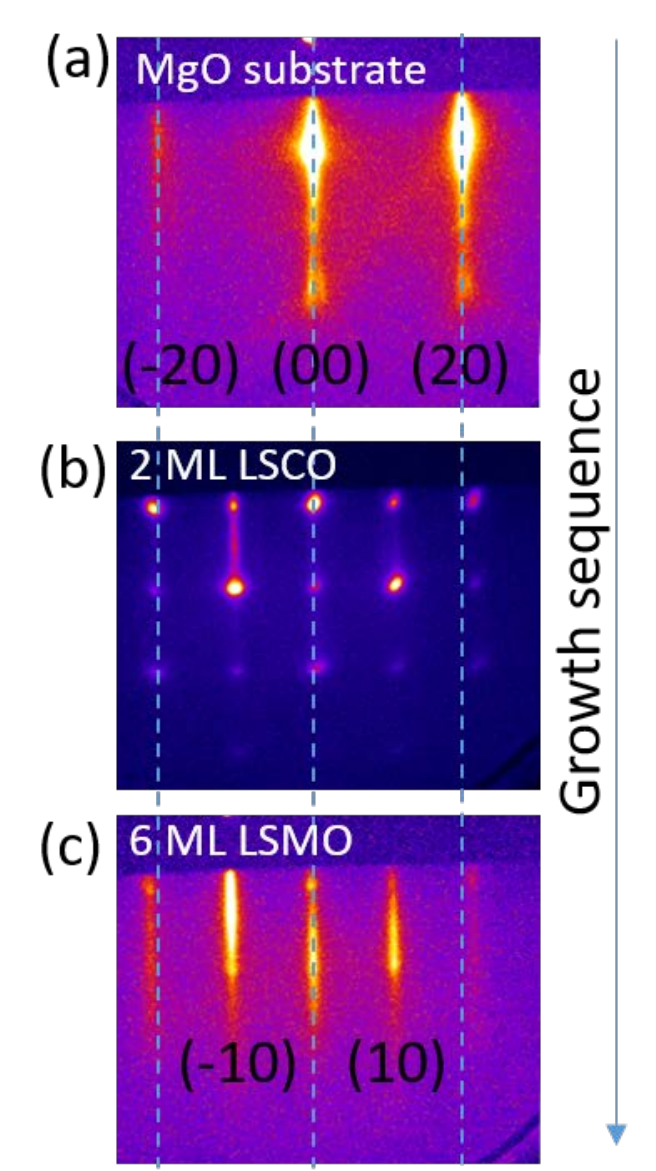}
    \caption{Reflection high energy diffraction (RHEED) pattern recorded along high symmetry zone axes during the growth of a 2 uc LSCO/6 uc LSMO  bilayer for (a) the initial MgO substrate (b) the first 2 uc of LSCO indicating 3D growth and for (c) 6 uc of LSMO indicating a transition to 2D growth. The shift of the (2 0) diffraction pattern of the film is indicative of a contraction of the in-plane lattice constant from 4.21 \AA{} for the MgO substrate to $\sim$ 3.87 \AA{} for the LSCO and LSMO layers. The vertical dashed lines indicate the position of the MgO (2 0) reflection.}
    \label{fig:RHEED}
\end{figure}


\section{Experimental procedure}

Trilayer 3 uc LSCO/ 3 uc LSMO/ 3uc LSCO films (referred to as (3/3/3)) and [2 uc LSCO/N uc LSMO]$_x$ superlattices (where x is the number of repeats) were fabricated by molecular beam epitaxy (MBE) on (001)-oriented MgO substrates. The LSMO layer thickness, N, was varied from 2-6 ucs and the superlattices were capped with 2 uc LSCO. Prior to growth, the MgO substrates were etched in buffered hydrofluoric acid and annealed in a tube furnace at 1200 $^o$C to achieve atomically flat surfaces. The films were grown at a substrate temperature of 950 $^o$C using an oxygen plasma source with an oxygen partial pressure of 3x10$^{-6}$ Torr. After growth, the films were cooled to room temperature at a rate of 5 $^o$C/min at the growth oxygen pressure to minimize the formation of oxygen vacancies. \textit{In-situ} reflection high energy electron diffraction (RHEED) was used to monitor the film thickness and crystallinity during growth.

To determine the strain states of the samples, reciprocal space maps of the 3/3/3 trilayer sample were measured at room temperature the 33ID beamline at the Advanced Photon Source with a Pilatus 100K pixel detector.\cite{Schleputz2005ImprovedDetector} Specular diffraction scans of the film and substrate Bragg peaks for the [2 LSCO/4 LSMO]$_6$ and [2 LSCO/6 LSMO]$_4$ superlattices were measured using a Rigaku Smartlab diffractometer equipped with a Ge(220) double bounce monochromator.

Temperature and magnetic field-dependent measurements of the magnetization of the samples were performed by superconducting quantum interference device (SQUID) magnetometry using a Quantum Design MPMS 3 system. The temperature-dependent magnetization curves were measured on warming up the sample with an applied 0.1 T magnetic field after field cooling in an 1 T magnetic field. The magnetization measurements were performed with the applied magnetic field oriented either parallel to the film [100] axis (in-plane) or to the film [001] axis (out-of-plane).

\section{Results and Discussion}
\subsection{X-ray diffraction measurements}

The evolution of the RHEED pattern during the growth of the first LSCO and LSMO layers is shown in Fig. \ref{fig:RHEED}. A transition in the RHEED spectra from the diffraction pattern for the MgO substrate surface in Fig. \ref{fig:RHEED}(a) to 3D spots (Fig. \ref{fig:RHEED}(b)) after deposition of 2 uc of LSCO is indicative of initial island growth and surface roughening. Additionally, an increase in the spacing of the RHEED streaks indicates a decrease in the in-plane lattice constant from 4.21 \AA{} for the MgO surface to $\sim$3.87 $\pm0.02$ \AA{} for the LSCO adlayer. On deposition of 6 uc of LSMO, the RHEED pattern transitions from the 3D pattern to 2D streaks shown in Fig. \ref{fig:RHEED}(c) indicative of a smoothening of the film surface. The initial roughening is attributed to a 3D Volmer-Weber island growth to relax the large film-substrate lattice mismatch.\cite{floro2001dynamic}

\begin{figure}[htbp]
    \centering
    \includegraphics[width=3.5in]{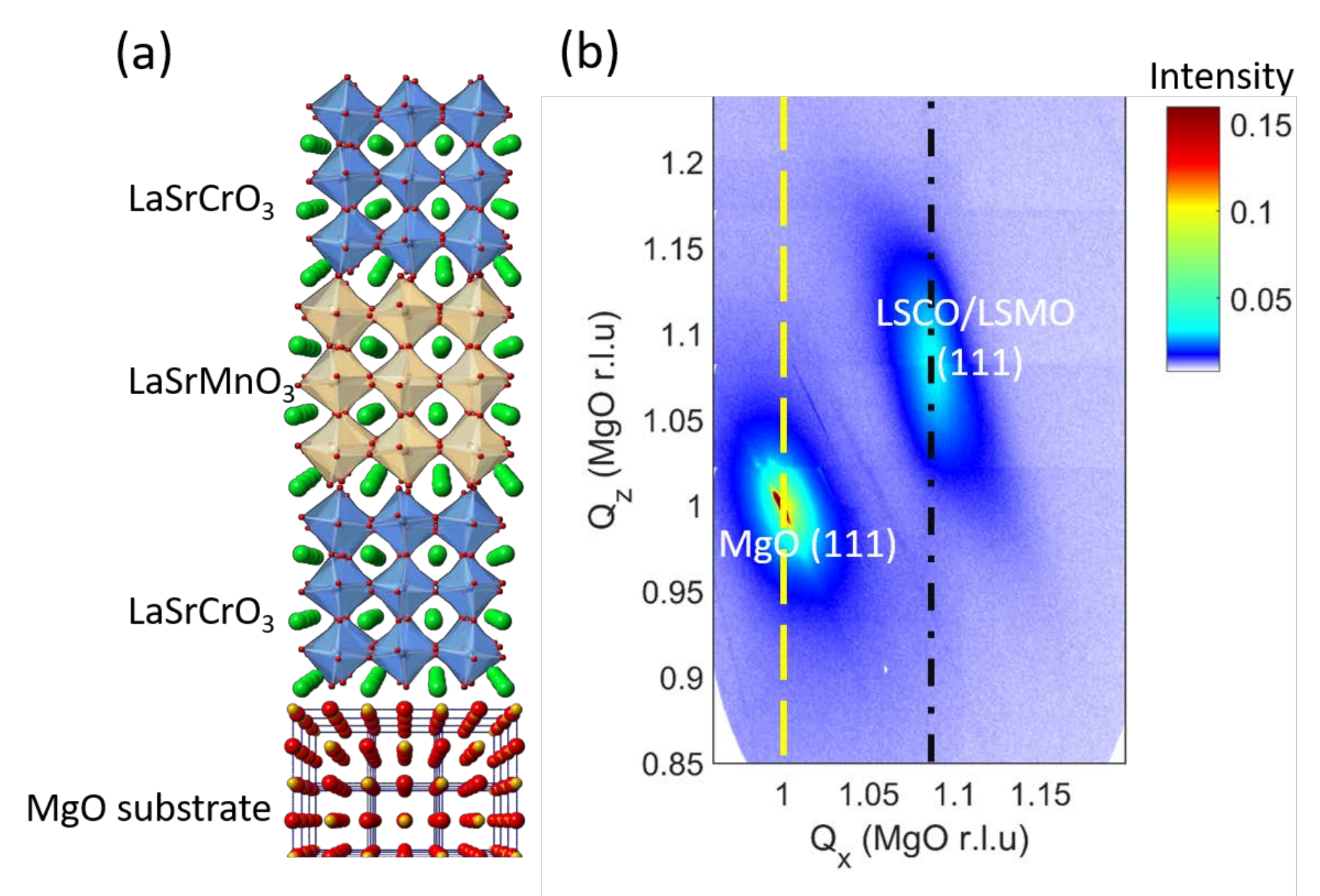}
    \caption{(a) Schematic of 3 uc LaSrCrO$_3$/ 3 uc LaSrMnO$_3$/ 3 uc LaSrCrO$_3$ heterostructure grown on (001)-oriented MgO by molecular beam epitaxy. (b) The reciprocal space map around the film and substrate (111) Bragg peaks indicate film strain relaxation. The yellow and black vertical dashed lines indicate the in-plane reciprocal vector corresponding to bulk MgO and LSMO respectively. }
    \label{fig:rsm}
\end{figure}

To confirm the relaxation of strain in the LSCO/LSMO layers, reciprocal space maps were measured at the 33ID beamline at the Advanced Photon Source with an incident photon energy of 15.5 keV. Fig. \ref{fig:rsm} shows the reciprocal space map around the LSMO/LSCO  film (111) and the MgO (111) Bragg peak for a trilayer 3 uc LSCO/ 3uc LSMO/ 3uc LSCO sample on MgO. The location of the film Bragg peak corresponds to a relaxed pseudo-cubic lattice constant of a=b=c=3.87 $\pm$ 0.01 \AA{}.

\begin{figure}[t!]
    \centering
    \includegraphics[width=3.5in]{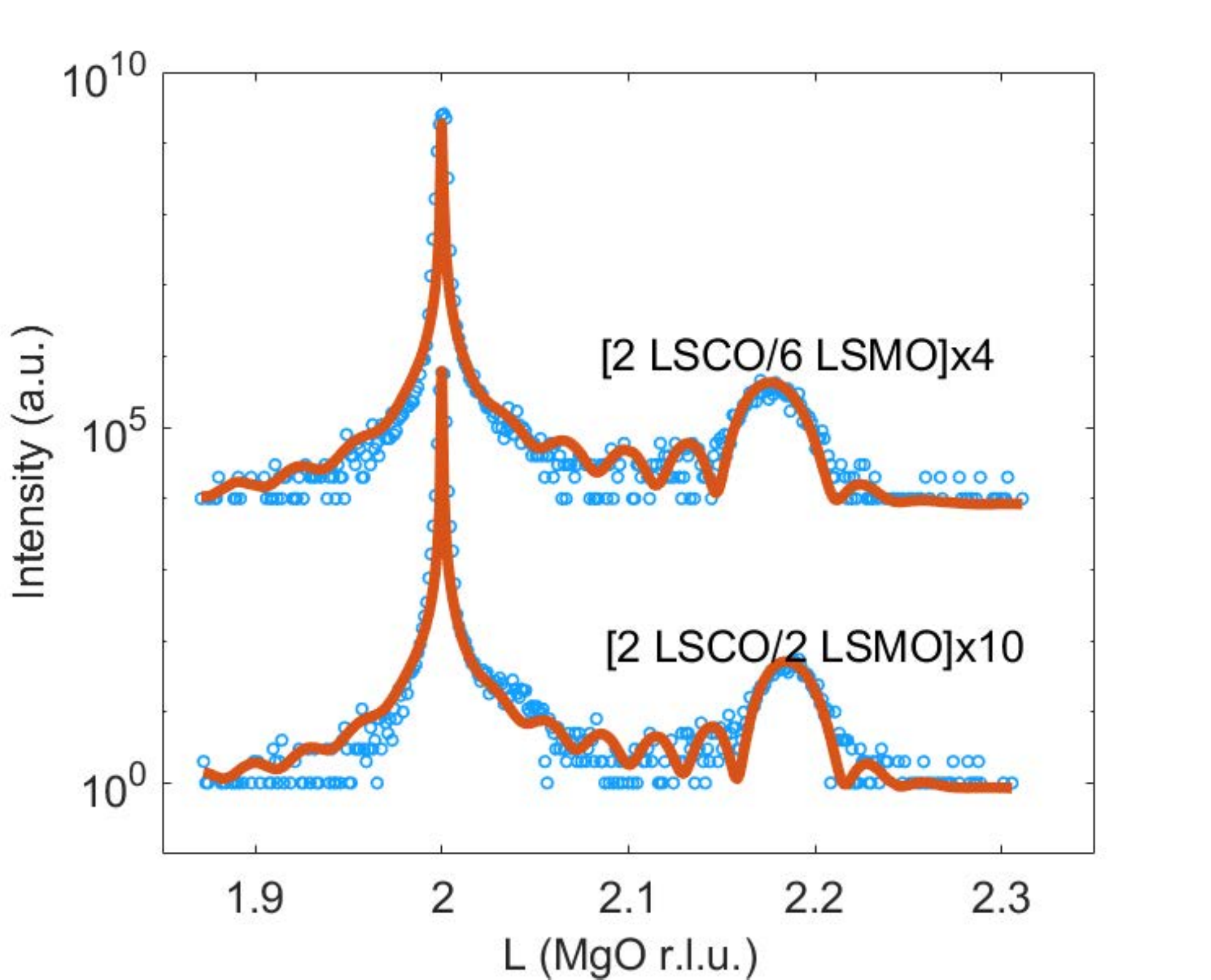}
    \caption{Measured (circles) and simulated (solid lines) specular diffraction of [2 LSCO/2 LSMO]$_{10}$ superlattice and [2 LSCO/6 LSMO]$_4$ superlattice on MgO substrate. The diffraction spectra are offset for clarity. L is the reciprocal lattice unit of the MgO substrate where 1 r.l.u.=$(1/4.212)$\AA{}$^{-1}$. }
    \label{fig:xrd}
\end{figure}

The out-of-plane lattice constants of the superlattices are determined from specular diffraction scans around the film and substrate (002) Bragg peaks. The (002) Bragg peaks for [2 uc LSCO/4 uc LSMO]$_6$ and [2 uc LSCO/6 uc LSMO]$_4$ superlattices on MgO substrates are shown in Fig. \ref{fig:xrd}. The fits to the measured data are obtained using the GenX X-ray analysis program.\cite{Bjorck2007GenX:Evolution} The average out-of-plane lattice constant determined from the fits for the LSMO and LSCO layers are  3.868$\pm 0.005$ and 3.862$\pm 0.005$ \AA{}  respectively. The calculated lattice parameters obtain from the fit is very close to the bulk LSMO lattice constant which further supports the lattice relaxation in LSCO/LSMO heterostructures on MgO.\cite{lebedev2003structure}

\subsection{Magnetization measurements}

\begin{figure*}[ht]
    \centering
    \includegraphics[width=6.5in]{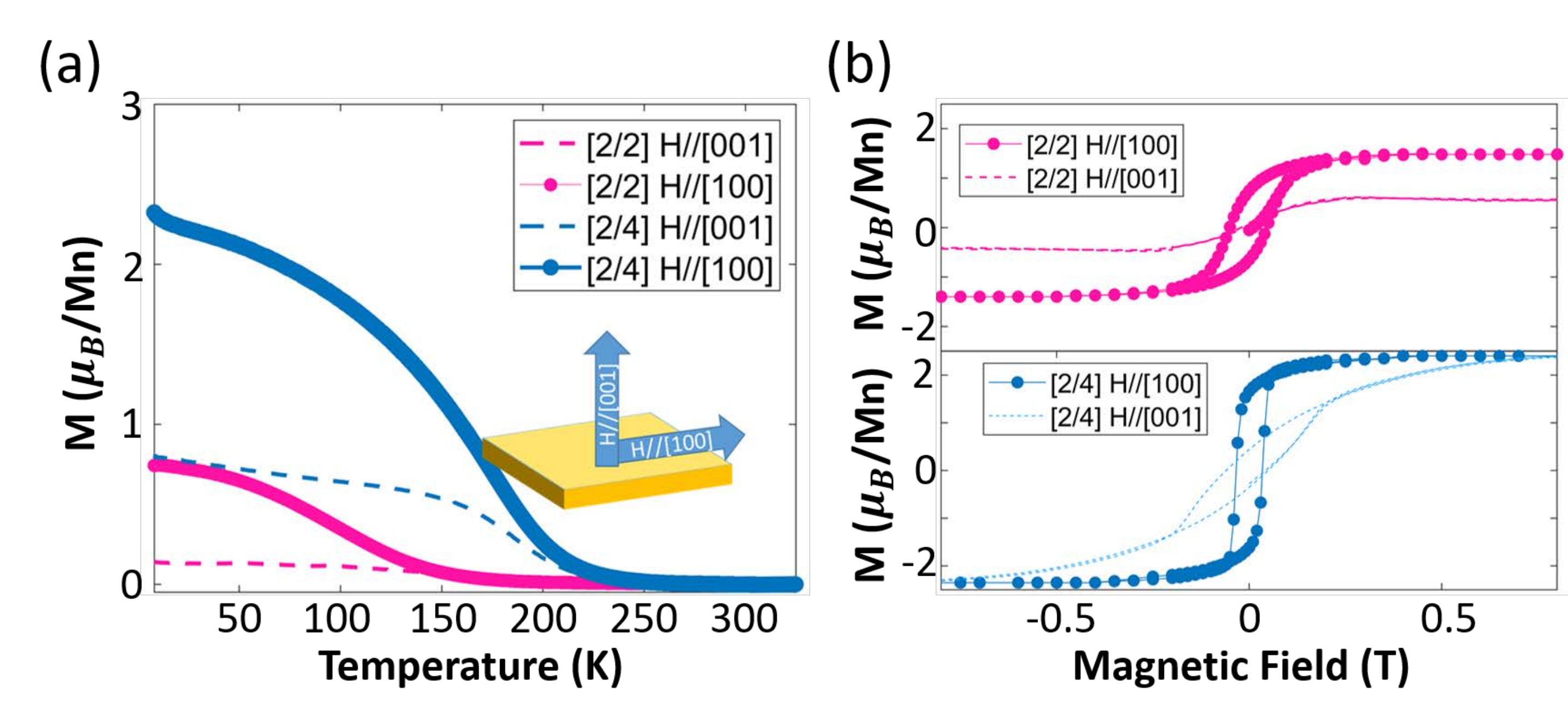}
    \caption{(a) Temperature-dependent in-plane and out-of-plane magnetization and (b) Field-dependent magnetization curves at 10 K for [2 LSCO/ 2 LSMO]$_{10}$ and [2 LSCO/ 4 LSMO]$_6$ superlattices grown on (001)-oriented MgO by molecular beam epitaxy.}
    \label{fig:squid}
\end{figure*}

To determine the effect of the lattice relaxation on the magnetic properties of the heterostructures, magnetization as a function of applied magnetic field and temperature were measured by SQUID magnetometry. The magnetization as a function of magnetic field and temperature for the [2 LSCO/2 LSMO]$_{10}$ and [2 LSCO/4 LSMO]$_6$  superlattices on MgO are shown in Fig. \ref{fig:squid}(a) and Fig. \ref{fig:squid}(b), respectively with the magnetic field oriented either in-plane or out-of plane as shown in the inset in Fig. \ref{fig:squid}(a). For both samples, the magnetization is significantly suppressed in the out-of-plane direction compared to the in-plane magnetization. This is indicative of an in-plane magnetic easy axis. For the [2 LSCO/2 LSMO]$_{10}$ sample, the out-of-plane magnetization does not reach the saturation value for an applied magnetic field of 1 T. Since the films are relaxed, the strong magnetic anisotropy is attributed to the LSMO/LSCO interfacial magnetic exchange interactions.

Furthermore, the observation of of ferromagnetism in the heterostructure with LSMO thickness of 2 uc is attributed to the removal of magnetic dead layers when using LSCO as spacer due to the interfacial structural coupling and the anti-ferromagnetic coupling between Cr and Mn across the LSCO/LSMO interface.\cite{Koohfar2019} The measured magnetic moments for the relaxed heterostructures on LSMO are expected to be close or identical to analogous heterostructures grown on LSAT where the lattice mismatch is small (0.2\%).\cite{koohfar2020effect} The magnetic moments for the [2 LSCO/2 LSMO]$_{10}$ and [2 LSCO/4 LSMO]$_6$ heterostructures on LSAT at 10 K are 1.5 $\mu_B/Mn$ and 2.8 $\mu_B/Mn$ respectively. The corresponding moments for the [2 LSCO/2 LSMO]$_{10}$ and [2 LSCO/4 LSMO]$_6$ on MgO are 1.4 and 2.4 $\mu_B/Mn$ respectively. The increase in the magnetization per Mn with increasing LSMO thickness is attributed to the contribution to the total magnetization of the Cr spins aligned anti-parallel to the applied magnetic field.\cite{Koohfar2019, koohfar2020effect}  





To quantify the magnetic anisotropy (MA) between the magnetic easy and hard axis, we calculate the magnitude of the effective anisotropy constant, $K_{eff}$, by the area enclosed between the two magnetization curves for the in-plane and out-of-plane applied magnetic fields shown in Fig. \ref{fig:squid}(b) for the [2 LSCO/4 LSMO]$_6$ heterostructure.\cite{johnson1996magnetic, yi2017tuning} $K_{eff}$ for the [2 LSCO/4 LSMO]$_6$ configuration is approximately $2\times 10^{5}$ ergs/cm$^3$ which is comparable to that of strained LSMO.\cite{suzuki1997role} This measured anisotropy for the relaxed heterostructure is unexpected since the magnetocrystalline anisotropy for bulk LSMO is 1 order of magnitude lower.\cite{suzuki1998magnetic}

An interface-induced magnet anisotropy has been reported for SrIrO$_3$/La$_{1-x}$Sr$_x$MnO$_3$ superlattices where an increase in the perpendicular magnetic anisotropy (PMA) is correlated with an increase in oxygen octahedral rotations about an in-plane axis as a function of the Sr content.\cite{Wu2017Interface-inducedFilms} For the relaxed LSCO/LSMO heterostructures, the in-plane and out-of-plane rotations are expected to be bulk-like and equivalent along the orthogonal pseudo-cubic axes. Hence, the octahedral rotations may be ruled out as the origin of the MA. A more probable origin of the MA is the asymmetry between the the Mn-O-Cr exchange across the interface and the in-plane Mn-O-Mn exchange interactions. Song \textit{et. al.}  show that the ferromagnetic interfacial Mn-O-Co exchange and charge transfer are associated with a PMA in LaSrCoO$_3$/LaSrMnO$_3$ bilayers.\cite{song2020electric} By tuning the Co charge states by ionic-liquid gating, an antiferromagnetic Mn-O-Co is stabilized leading to a reorientation of the magnetic easy axis in the in-plane direction.\cite{song2020electric} The Mn-O-Cr exchange has previously been found to be antiferromagnetic and independent of the substrate-induced epitaxial strain.\cite{koohfar2020effect}  Hence, we conclude that the in-plane easy axis for the LSCO/LSMO heterostructures is driven by the anti-ferromagnetic interfacial superexchange.



\section{Conclusion}
In conclusion we have investigated the effect of LSCO spacer layers on the magnetic properties of thin LSMO layers grown on (001)-MgO substrates. The LSCO/LSMO heterostructures are found to be relaxed as a result of the large lattice mismatch with the MgO substrate with strain relaxation occurring within 2-3 unit cells. The magnetic easy axis of the LSCO/LSMO heterostructures is found to lie in the plane of the film due to the antiferromagnetic Cr-O-Mn exchange across the LSMO-LSCO interface. Our work demonstrates the critical role interfacial interactions play in modulating the magnetic states of transition metal oxide heterostructures.

\begin{acknowledgments}
 S.K., Y.O. and D.P.K. acknowledge financial support by the US National Science Foundation under Grant No. NSF DMR-1751455. This work was performed in part at the Analytical Instrumentation Facility at North Carolina State University, which is supported by the State of North Carolina and the National Science Foundation (award number ECCS-1542015). This work made use of instrumentation at AIF acquired with support from the National Science Foundation (DMR-1726294). The AIF is a member of the North Carolina Research Triangle Nanotechnology Network (RTNN), a site in the National Nanotechnology Coordinated Infrastructure (NNCI). The authors acknowledge use of the SQUID and PPMS facility in the Department of Materials Science and Engineering at North Carolina State University. Use of the Advanced Photon Source was supported by the U.S. Department of Energy, Office of Science, Office of Basic Energy Sciences, under Contract No. DE-AC02-06CH11357.
\end{acknowledgments}

\section*{Data Availability}
The data that support the findings of this study are available from the corresponding author upon reasonable request.


\bibliography{refs}

\end{document}